How can we improve problem solving in undergraduate biology? Applying lessons from 30 years of physics education research

A.-M. Hoskinson[1], M.D. Caballero[2], J.K. Knight[3]

[1]Department of Ecology and Evolutionary Biology, University of Colorado Boulder
[2]Department of Physics, University of Colorado Boulder
[3]Department of Molecular, Cellular, and Developmental Biology, University of Colorado Boulder

Abstract. If students are to successfully grapple with authentic, complex biological problems as scientists and citizens, they need practice solving such problems during their undergraduate years. Physics education researchers have investigated student problem solving for the last three decades. Although physics and biology problems differ in structure and content, the instructional purposes align closely: explaining patterns and processes in the natural world and making predictions about physical and biological systems. In this paper, we discuss how research-supported approaches developed by physics education researchers can be adopted by biologists to enhance student problem-solving skills. First, we compare the problems that biology students are typically asked to solve with authentic, complex problems. We then describe the development of research-validated physics curricula emphasizing process skills in problem solving. We show that solving authentic, complex biology problems requires many of the same skills that practicing physicists and biologists use in representing problems, seeking relationships, making predictions, and verifying or checking solutions. We assert that acquiring these skills can help biology students become competent problem solvers. Finally, we propose how biology scholars can apply lessons from physics education in their classrooms and inspire new studies in biology education research.

How can we improve problem solving in undergraduate biology? Applying lessons from 30 years of physics education research

A.-M. Hoskinson[1], M. D. Caballero[2], J.K. Knight[3]
[1]Department of Ecology and Evolutionary Biology, University of Colorado Boulder
[2]Department of Physics, University of Colorado Boulder
[3]Department of Molecular, Cellular, and Developmental Biology, University of Colorado Boulder

## I. Introduction

In 2009, over 85,000 students earned undergraduate biology degrees at colleges and universities in the United States. These students represent the second largest single population of science majors at American colleges and universities, trailing only psychology (National Science Board 2012). Furthermore, in 2009, of the approximately 74,000 students enrolling for the first time in science graduate programs, 40% were in biology or medical-related fields (National Science Board 2012), where a strong foundation in solving problems will facilitate their success (NAS 2011). Whether or not students graduating with a bachelor's degree in biology go on to use biology in their professions, they will undoubtedly confront complex problems as citizens: making medical decisions, engaging in conservation planning, and understanding climate change. Thus, no matter what their future careers, it is paramount that students in the biological sciences become capable of grappling with complex problems (AAAS 2011).

Based on introductory science enrollments, biology is likely the undergraduate discipline where most students are first exposed to *scientific process skills* such as developing hypotheses, interpreting data, crafting evidence-based arguments, and using and evaluating models of systems (National Science Board 2012). These process skills are called *scientific practices* in much of the physics education literature and in the forthcoming K-12 Next Generation Science Standards (National Research Council 2012b). Helping students learn such skills facilitates their development into competent problem solvers, and exposes them to how scientists think about complex ideas in their disciplines (Dunbar 2000, Taconis et al. 2001). If students are to be successful in such problem solving later in life, they need to have extensively practiced such skills in their undergraduate science classes (AAAS 2011, Jonassen 2011, NAS 2011).

 "Problem solving" appears regularly in the biology education research (BER) literature, and there is near-universal agreement that problem solving is a valuable skill for biology students to learn and practice (AAAS 2011, NAS 2011). Research about defining relevant biology problems, how students solve biology problems, and what conceptual knowledge they employ while solving biology problems is still in formative stages. Furthermore, developing effective biology curricula that incorporate problem solving will depend on the outcome of such research (AAAS 2011, National Research Council 2012a). To help direct the future of problem solving in biology, BER scholars can gain insight from the work in other discipline-based education research (DBER) fields – namely, physics education research (PER).

PER scholars have investigated problem solving in physics for the last three decades (Hsu et al. 2004), including defining problems (Heller and Reif 1984), investigating how students solve problems (Larkin et al. 1980), the role of conceptual knowledge in solving problems (Reif 2008), and developing pedagogical and curricular tools that facilitate deeper learning (Heller and Hollabaugh 1992, Mestre et al. 1993, Pawl et al. 2009). Historically, such research has concentrated on developing students' abilities to grapple with end-of-chapter physics problems by focusing student attention first on analysis of the underlying physics principles (i.e., employing *conceptual knowledge*). This work has spanned levels of instruction from middle and high school courses to upper-division undergraduate courses, has helped to improve student conceptual knowledge (Kohlmyer et al. 2009), and, in some cases, has resulted in the development of research-based teaching practices in problem solving (Heller and Hollabaugh 1992, Heller et al. 1992). More recently, the PER community has found that by emphasizing scientific practices (*process skill*s in BER) in both curricula and pedagogy, gains in problem-solving competency can be achieved (Hestenes 2000). The PER community has broadened the definition of problem solving to include employing aspects of professional science: for example, using conceptual knowledge to design experiments, develop and test models, and critique scientific information (Meltzer and Thornton 2012).

In this paper, we discuss research-based approaches, first developed by PER scholars, that can be used by biologists to increase student problem-solving mastery, and to inform future research into problem solving in biology. By contrasting authentic, complex problems with the kinds of problems that biology students are typically asked to solve (Section II), we illuminate a significant gap between existing curricular methods and the professional practices of biologists. We summarize the work done by PER scholars that narrowed a similar gap between physics education and the practice of physics in Section III. We conclude with a discussion of how biology educators can put lessons from PER into practice, and suggest new directions for research into biology problem-solving theory and practice (Section IV).

## II. Undergraduate biology and complex problems: A coursework-practice gap

There are many ways to describe problem-solving processes. Here, we focus on problems and problem-solving strategies. When biology researchers investigate a topic, they engage in a variety of scientific practices that are routine for scientists, but often difficult for students: these are authentic and complex problems (Chi 2005, AAAS 2011). Problem solving activities designed for students can be characterized along two axes: by the *content* and *structure* of the problems to be discovered and solved; and by the *process skills* or *scientific practices* that problem solvers must engage in to achieve a solution (Table 1) (Fischer and Greiff 2012).

The phrase "authentic problem" has multiple meanings in BER, including "real-world" problems (Jonassen 2011, Gormally et al. 2012), problems with personal and social relevance (Hanauer et al. 2006, Wenglinsky and Silverstein 2007), and problems with multiple possible solutions (Steen 2005, AAAS 2011, Brownell et al. 2011). Likewise, "complexity" has both colloquial and discipline-specific meanings. However there is good alignment about what constitutes a "complex problem," whether in physics or biology (Goldenfeld 1999, Fischer and Greiff 2012). First, complex problems often **explore systems and phenomena that are dynamic, non-linear, stochastic, and/or emergent** (e.g., interactions among gene sequences, probabilistic mutations, and chromosomal segregation in inheritance; interactions among animals, plant, climate, nutrients, etc. in a watershed; the relevant model needed to design a working catapult; Table 1:D-F). Complex problems have **multiple elements, features, or variables of the system that must be considered** (i.e., the elements of inheritance include gene sequences, mutation sites, enzymes, and chromosomes; the elements of an ecosystem include organisms, environment, nutrients; the elements of dynamics include force, acceleration, velocity, position). The elements have **variable relationships with one another**, not necessarily defined in the problem description (i.e., the relationships between genes and their expression is moderated by probabilistic mutation and gene regulation; the relationship between elements of urban streams is driven by intra- and inter-annual climate variation; the relationships governing motion are Newton's Second Law and kinematics).

Complex problems can also be considered in terms of how the problem solver must interact with the problems by using particular skills and practices. Unlike routine problems, or *exercises* (Smith and Good 1984), complex problems cannot be solved merely by recalling facts from memory, or by using a simple algorithm. Instead, complex problems require that **problem solvers engage in a wider variety of scientific practices** – processes and skills specifically defined as complex problem solving in the literature (*CPS*; Frensch and Funke 1995, Jacobson 2001, Fischer and Greiff 2012). This second aspect of problem solving emphasizes the levels of cognitive functioning or expert-like skills necessary for solving the problem. Problems that involve scientific practices such as analyzing data, evaluating outcomes, employing conceptual knowledge, and designing experiments require higher levels of cognitive functioning than exercises that require merely recalling simple facts (Bloom 1956).

In addition to the higher levels of cognitive functioning needed, CPS requires that **problem solvers characterize problems in expert-like ways**. Expert problem solvers and novices are known to differ both in their characterizations of complex problems (Chi and Slotta 1994, Jacobson 2001) and in the processes, methods, and skills each group employs to solve them. Novices are more likely to attribute external control to complex systems and to focus on stepwise solutions. Experts acknowledge attributes of complex systems, such as emergence and stochasticity (Chi et al. 1981, Chi 2008). Experts also seek multiple solution pathways (Taconis et al. 2001), use more sophisticated heuristics (Nicolson et al. 2002, Fischer and Greiff 2012), and employ cognitive flexibility and filtering (DeHaan 2009) to reduce both solution space and potential solution pathways (Table 1). Many biologists agree that students should learn to solve biology problems requiring complex problem-solving skills (Taconis et al. 2001, Jonassen 2011). Often, though, the problems students solve in biology classes are simple exercises in which the system is well understood, most or all variables and relationships are given, a solution path (algorithm) is given or known, and the solution or answer is pre-defined. Consider an example of an exercise many genetics students work through (Table 1:A). In this case, the system – chromosomal inheritance – is well understood. All of the variables are given, and there is one correct way to calculate the single correct answer. Similar examples abound in other domains of biology (Table 1:B), and in physics (Table 1:C).

Alternatively, consider the kinds of problems that scientifically literate citizens should be able to solve (Table 1: D-F). In these problems and many others, the understanding of the system is vague, or may not be shared among the problem solvers. This can lead to multiple, potentially competing ideas of what constitutes a solution or answer (Nicolson et al. 2002, Fischer and Greiff 2012). The representations of both problem elements and relationships among the elements are poor (Jonassen 2011, Fischer and Greiff 2012). As the number of elements and relationships increases, their interdependence makes simpler problem representations and paper-and-pencil solutions by single individuals less feasible. Consequently, cooperation among multiple people – population ecologists, environmental chemists, economists, politicians, citizens – may be required to adequately represent and solve the problem (Table 1: D, E). The need for communication among experts and non-experts may also be featured (Table 1: D, E).

To solve these authentic, complex problems, participants engage in a wider variety of scientific practices than when solving simple exercises. These kinds of problems are infrequently presented in most of undergraduate biology curricula; thus there remains a significant gap between the kind of problems students can solve and the kind of problems they should be able to solve.

## III. Narrowing the coursework-practice gap in physics

Over the last 30 years, physics education researchers have investigated how students learn to solve problems in physics courses through a variety of perspectives (McDermott and Redish 1999, Hsu et al. 2004). This work has lead to a substantial number of research-based curricular and pedagogical tools that improve student learning in physics (Meltzer and Thornton 2012). Despite significant strides made by PER, many researchers and instructors are just now beginning to make use of authentic, complex problems. There is still much work to be done in PER to describe what authentic and complex physics problems look like, how students engage in scientific practices when solving such problems, and how students who learn to solve authentic and complex physics problems perform on more traditional assessments of problem solving.

Early work in PER highlighted the substantial differences between how physics students and experts think about the nature of science, including how their knowledge of physics principles is organized and employed, and what practices they employ to solve physics problems (Chi et al. 1981, Chi et al. 1989, Reif 2008). Representative problems were typical exercises. This research illuminated how students could successfully solve exercises without conceptual knowledge of the underlying physics. Unlike novice students, physics experts rarely begin solving problems by using mathematical equations. Just as in the biological sciences, physics experts often think of a few governing principles and heuristics, then construct models to make sense of physical phenomena. Often, they begin with their conceptual knowledge of the problem, and then develop it to include mathematical representations. Further refinement and mathematical manipulations lead to appropriate expressions of the problem (Reif 2008).

Novices, on the other hand, often think of physics as a loose collection of ideas and equations with few or no connections among them (Chi et al. 1981). When solving exercises, students rarely employ their conceptual knowledge, preferring instead to hunt for equations that contain all the elements (e.g., velocity, acceleration, mass) given in the problem statement (Chi et al. 1989, Reif 2008). Practices such as equation hunting emphasize memorization strategies that facilitate students' strong performance on traditional exams, which are typically constructed from exercises. Traditional exams tend to reward finding a single solution to an exercise ("answer making") rather than demonstrating a deep understanding of principles and methods ("sense making;" McDermott and Redish 1999, Meltzer and Thornton 2012).

Early work in PER demonstrated that, in addition to applying conceptual knowledge to solve exercises, students needed to learn to construct a variety of representations of physical systems, to coordinate between those representations, and to execute the necessary mathematics to successfully solve problems in physics. While this early work did not seek to define the authenticity or complexity of problems *per se*, it did provide the necessary foundation for researchers and instructors to develop transformative teaching methods. By emphasizing a number of scientific practices that were not present in traditional lecture courses, teaching methods emphasized authentic problems. For example, when a standard exercise is reframed as a design activity (Table 1: F), students must confront how a problem is defined, how a model can be constructed, and how variables can be reduced so the problem can be solved using the elementary physics and mathematics they are learning.

Building on the work of Chi, McDermott, Reif and others, Heller and colleagues formalized the tasks needed for solving typical exercises into a coherent framework (Heller and Heller 1995). In one instantiation of this framework, Heller and colleagues developed a suite of authentic problems not typically represented in physics curricula. While these problems were not necessarily complex (they still resolved to a single solution), they were authentic (*context-rich* in Heller and Heller; real world scenarios, often with personal relevance). These problems provided students with real-world scenarios (" you are a stunt driver jumping a series of buses") in which decisions had to be made about particular variables (e.g., mass of car + driver, initial velocity after leaving ramp) to facilitate a solution (e.g., landing safely). To solve these problems, students employed a stepwise framework: (1) focus on the problem, (2) describe the physics, (3) plan the solution, (4) execute the plan, and (5) evaluate the answer. This framework makes explicit use of conceptual knowledge in steps 1 and 2 and connects that knowledge directly to representing the problem mathematically in step 3 (Heller and Heller 1999).

In addition to using problems that were context-rich, Heller and Heller also made use of cooperative groups, which involved students in all aspects of the problem-solving process, offered opportunities for students to critique their peers' problem-solving practices, and facilitated the use of more challenging problems (Heller and Hollabaugh 1992, Heller et al. 1992). Explicit teaching of problem-solving practices in this way (e.g., activating and employing conceptual knowledge early in the problem-solving process) helped students develop such skills more quickly, and these students performed better on qualitative exam questions than traditionally-taught students (Foster 2000). Engaging students with context-rich problems in a cooperative group format is but one of a number of attempts by the PER community to develop students' abilities to solve problems in the traditional sense. In another example, O'Kuma and colleagues (2000) developed a series of ranking tasks to elicit student ideas about physical systems rather than memorized responses. The structure and content of such problems were not particularly authentic (e.g., limited personal or social relevance), but these tasks increased the complexity of usual course activities. Students attended to multiple elements in a single scenario (e.g., mass and velocity of a car) while considering the relationships within groups of elements and among scenarios (e.g., ranking car crash scenarios in order of the force experienced). Such tasks promoted both students' conceptual knowledge and their traditional problem-solving skills (O'Kuma et al. 2000). These tasks have been broadened to include a variety of alternative problem types and to engage students in more authentic scientific practice. Many of the newer activities require students to utilize their conceptual knowledge to explain their solutions to peers or critique the solution offered by others (Hieggelke et al. 2006a, Hieggelke et al. 2006b). Others have leveraged alternative representations (verbal descriptions, diagrams, graphs, and equations) to focus novice problem-solvers on performing a conceptual analysis first by coordinating between the different representations (Van Heuvelen and Maloney 1999, Van Heuvelen and Zou 2001).

More recently, PER has begun work at the upper-division undergraduate level, where course goals and the activities pursuing those goals are significantly more complex (e.g., using and connecting more sophisticated mathematical and physical ideas). For typical upper-division problems, students often grapple with complex systems in addition to considering how multiple elements and their relationships facilitate a solution (e.g., using the calculus of variations to determine Snell's law). Moreover, students in upper-division physics become acculturated to the scientific practices of professional physicists (e.g., developing and using models, gathering evidence, evaluating outcomes). Despite the increasing complexity of these problems, many upper-division physics problems are still inauthentic (idealized models of systems with limited personal relevance), and upper-division students struggle to solve these problems (Pepper et al. 2012). To uncover why students struggle, Caballero and colleagues developed a framework that has been used to analyze how students solve problems in upper-division physics courses (Caballero et al. 2013, Wilcox et al. 2013). Their work investigates how students blend their conceptual knowledge with problem solving practices to achieve solutions. Preliminary findings show that these physics students' primary difficulties include constructing precise models and evaluating the appropriateness of solutions. These scientific practices are two that professional physicists use daily, and students' difficulties with these practices must be addressed to strengthen their professional preparation.

While the previous examples of research-based tools emphasized, to varying degrees, either authentic or complex problems, physics teaching that focuses on developing students' ability to engage in scientific practice tend to use both. In physics, scientific practices include constructing and evaluating models, designing and executing experiments, and engaging in argumentation based on evidence. Scientific practices underpin what it means to engage in complex problem solving; in fact, these are the practices that professional scientists use to solve challenging problems in their own work (National Research Council 2012a). Through this lens, recent reforms in introductory physics have broadened the traditional definition of problem solving to include engaging in the practices of professional science. Consider again the design problem posed in Table 1:F: such a problem engages students in the practice of science, and thus can be characterized as an authentic and complex problem.

There are a number of curricular examples that emphasize scientific practices, and, hence, this broader definition of problem solving, such as Workshop Physics (Laws 1991, Etkina and Van Heuvelen 2007). Modeling Instruction (Hestenes 1987) is another approach gaining wider acceptance both in PER and the broader physics community. Modeling Instruction is worth describing in some detail because it has been implemented in both high school (Hestenes et al. 1995) and university (Brewe 2008) settings. Moreover, approximately 10% of the nation's high school physics teachers have had some formal training in Modeling Instruction.

Modeling Instruction employs a theoretical framework (the modeling cycle) around which student activities are organized. Students engage in open-ended experimental and theoretical procedures while making real-world observations, then propose possible measurements that can help describe observed patterns. From these measurements, students observe trends and patterns that help to inform their development of representative models. Then, they test their models against additional observations to evaluate their model's capacity to explain the observed phenomena. Students subsequently repeat this observation-development-evaluation cycle for new phenomena. Through this cycle, students discover the necessary elements and their relationships that describe the observed system. While the physics and mathematics are not particularly complex, students utilize several scientific practices during a single cycle: developing and using models, finding patterns, testing hypotheses. The Modeling Instruction curriculum is demonstrably effective in teaching physics concepts (Hestenes 2000), in developing students' self-efficacy (Sawtelle et al. 2010), in enriching their notions about the nature of science (Brewe et al. 2009), and in improving students' traditional problem-solving competence (Malone 2008). Both Workshop Physics and ISLE are similarly organized, with slightly differing pedagogies but equally effective results (Redish and Steinberg 1999, Etkina 2006). Physics courses that emphasize scientific practices will likely serve students well in their future coursework and beyond.

While many teaching methods have been developed by the PER community, effectiveness is often evaluated using end-of-course conceptual assessments. Concept inventories (assessments) measure whether students can show evidence of deeper learning from particular instructional strategies (Hestenes et al. 1992, Beichner 1994, Ding et al. 2006). Through measuring student-learning gains, these assessments have demonstrated the benefits of using active engagement scientific practices in teaching (Hake 1998, Hestenes 2000, Pollock and Finkelstein in press). Concept inventories have also been used to quantify the outcomes of introductory and upper-division physics courses (Kohlmyer et al. 2009, Caballero et al. 2012, Chasteen et al. 2012). Despite their value, most physics concept inventories do not directly measure problem solving – even narrowly defined. Thus, they may or may not serve well as predictors of such skills in students. One instrument that begins to characterize students' problem-solving skills in exercises is the Mechanics Baseline Test (MBT; Hestenes and Wells 1992). However, this instrument cannot directly evaluate authentic scientific practice skills, because investigating how students employ scientific practices is not amenable to the MBT's multiple-choice format. Future work in PER is needed to assess how students employ scientific practices.

## IV. How can PER inform BER on problem solving?

Theoretically, biologists should be able to apply the approaches that physics education researchers have found successful in improving instruction and student learning, especially because some of the same scientific practices have already been identified as critical in biology (e.g., formulating hypotheses, gathering and evaluating evidence, and using evidence to construct arguments). However, biology education researchers are still working to define what it looks like for students to solve complex problems in biology. We focus below on ways in which the findings on problem solving in PER can specifically impact curriculum development and research on problem solving in biology.

### Problem solving in the biology classroom
New curricula for biology students should include ways for students to use scientific practices and process skills, and to engage in complex problem solving, as described above.  Currently, many problems that biology students are asked to solve are exercises. Consider an exercise that a biology student might encounter in an introductory genetics course: calculating inheritance probabilities of sex-linked traits. A problem on this topic could be presented as follows:

> "Suppose that hemophilia is an X-linked recessive trait. If a mother is a carrier for hemophilia, and the father is normal, what is the chance that their son will have hemophilia?"

To make the problem more complex and engage students in more authentic problem-solving practices, the prompt can emphasize the construction (design) of a possible pedigree, such as:

> "Generate a possible pedigree for three generations showing unaffected, affected, and carrier individuals in a pedigree for hemophilia. Share your pedigree with your neighbor. How are the two pedigrees different? Which is more likely to occur, given the history of hemophilia?"

This approach (as well as the other examples included in Table 1) transforms a typical exercise into a complex problem, inviting students to generate possible scenarios by applying their knowledge of a generalized system (chromosomal inheritance) to a specific application.

Such problems could also be presented to students as in-class concept "clicker" questions. In the Peer Instruction model originally proposed by Mazur (1997), multiple choice in-class questions can be used to foster discussion amongst students, so that they are engaged in a community of problem solvers. Clickers work well for implementing this active learning technique by both biologists and physicists because they are easy to incorporate, especially in large classes, and they can provide immediate feedback to both students and instructor (Wood 2004). Some may view multiple-choice clicker questions as limited, since the potential answers are defined; however, it is possible to write multiple-choice questions that require higher-order thinking and that require students to engage in complex problem solving (Crowe et al. 2008). Curricula that encourage the use of complex problem solving in class, as practices or as part of formative assessment, have the potential to foster complex problem-solving abilities in students (DeHaan 2009, Maskiewicz et al. 2012).

The observation-development-evaluation cycle of Modeling Instruction could also be adapted to biological problems and implemented in a variety of classroom settings. Some biology resources already exist that could be adapted for this purpose, such as case studies (NCCSTS 2012, ESA 2013) and problem-based scenarios (Norman and Schmidt 2000). These examples begin with authentic, catchy stories to pique students' interest, usually describing observations from the viewpoint of science students rather than expert scientists (personal relevance). Then, students develop experiments to test hypotheses, to explore the sensitivity of a model to changes in variables or parameters, or to suggest and justify what further measurements they would make (authentic scientific practice). The final step in case-based or problem-based curricula is to evaluate evidence against observations, and to construct arguments for a resolution to a problem or dilemma. This kind of curricular path, then, builds in higher-order process skills that support student problem solving.

Following the work of PER scholars who introduced the importance of measuring conceptual understanding and sense-making with concept inventories (e.g. Hestenes et al. 1992), biologists have recently developed a variety of such assessments in different sub-disciplines of biology (D'Avanzo 2008). Not all concept inventories have been developed with attention to evidence of validity and reliability (Huffman and Heller 1995), and few directly measure problem solving or critical thinking skills (Smith and Tanner 2010), as also described above. Nonetheless, several of the most recently developed biology concept inventories/assessments have demonstrated utility in uncovering concepts for which students have persistent difficulties (Garvin-Doxas and Klymkowsky 2008, Parker et al. 2012, Smith and Knight 2012). In addition, several recently-developed assessments are devoted at least in part to specifically measuring problem solving (i.e. Diagnostic Question Clusters; Parker et al. 2012), critical thinking (Bissell and Lemons 2006), and basic process skills (Gormally et al. 2012). In fact, assessing how students engage in scientific practices is important to all DBER communities (National Research Council 2012a). Student performance on these kinds of tools, coupled with the results of formative assessments devoted to problem solving, may be useful in further informing curricular change.

**Research**

Though gaps between theory and practice remain, these gaps also suggest rich opportunities for research. For instance, as described in Section II, some BER scholars have begun to investigate how student conceptions of biological principles and processes differ from experts. One approach to solidify our understanding of biological problem solving would be to ask what process skills novices tend to employ when problem solving, compared to skills that experts typically employ. Similar investigations were helpful in framing the development of research-based instructional materials for introductory physics courses (McDermott and Shaffer 2002). Understanding the processes that expert biologists use to solve problems could be used to help teach the scaffolding of both content knowledge and process skills to students. Process skills may also be an important mechanism linking problem-solving activity and conceptual knowledge.

Although there exist ways of measuring some aspects of problem solving, and other methods to assess conceptual knowledge, there are as yet few investigations of ways to measure conceptual knowledge using the processes of problem solving (Nehm 2010). Diagnostic question clustering (DQC; Parker et. al 2012) and Bissell and Lemon's (2006) methods are encouraging steps toward establishing and exploring these links. Work that explores the kinds of links between conceptual knowledge and sense making with actual problem solving competence should also focus on developing assessments that are validated and easy to deploy.

One important difference between physics and biology lies in how problems represent systems behavior. Physics problems tend to emphasize quantitative representations. Qualitative, conceptual, and pictorial representations are used extensively in physics, but, ultimately, the goal of many problems is to connect the physics to the mathematics to make predictions. Whether simple or complex, such problems require the problem-solver to use quantitative representations such as equations, graphs, or predictive models. While some biological problems are quantitative, many others rely upon different representations, including diagrammatic or pictorial representations such as those used to represent signaling pathways, biological cycles, or relationships among cycles (i.e. life, carbon, nutrients, populations). A few studies have investigated how students perceive and interpret the representations drawn by experts (Schönborn and Anderson 2009), and how such representations can introduce and perpetuate misconceptions (Catley and Novick 2008), but there is relatively little work on how students generate their own representations (Kose 2008), or how their representations impact their problem-solving skills.

Understanding and coordinating between representations is but one distinguishing feature and promising area of research into complex problem solving in biology. Other process skills may be just as necessary and important for biology students to learn. Metacognition is widely recognized as important for learning (Tanner 2012), and it may be especially important in CPS for progress checking and for reconciling feedback among potential solutions, representations, and mental models. Decision-making as a process skill may also be fundamental to CPS (Nicolson et al. 2002, Fischer and Greiff 2012).

Although we have identified several structures and processes common to complex problem solving, this is by no means a comprehensive list, or an operational model for designing problems or assessing problem solving. The development of operational models of CPS processes would help target the behaviors and skills necessary for students to engage in solving authentic, complex problems in the classroom and in their lives. We live at a time when complex biological problems are not just the realm or responsibility of highly trained scientists. The people who are currently our students will need to engage along with us as citizens once they leave college. Thus, it serves all of us to engage in solving the kinds of problems that will emerge in our collective futures.

## Acknowledgements

The Science Education Initiative at the University of Colorado Boulder supported AMH and MDC. The authors thank B. Couch, J. Jackson, B. Zwickl, and three anonymous reviewers for comments that improved the manuscript.

Table 1. Attributes of problem solving (top row) characterized by features of the problem and skills required. Several examples of simple problems (exercises; A-C) and complex problems (D-F) illustrate the distinct features of simple and complex problems.

| | Example problem | Problem Features | | | Process Skills (Practices) | |
|---|---|---|---|---|---|---|
| | | Elements | Relationships | Solutions | Lower-order cognitive skills required | Higher-order cognitive skills required |
| A. Simple genetics exercise | Cystic fibrosis is an autosomal recessive disease. If two individuals, who are both carriers of the same cystic fibrosis mutation, have a child together, what is the probability that their child will be a carrier? | Alleles, Chromosome segregation, Probability | Deterministic probability | One | Recall facts about inheritance and probability Solve the equation | Turn a verbal representation into an equation |
| B. Simple ecology exercise | Given a current population size and intrinsic growth rate, predict a future population size | Population size, Growth rate | Deterministic | One | Recall facts about population growth Solve the equation | Turn a verbal representation into an equation |
| C. Simple physics exercise | An object is thrown horizontally with a speed of 20m/s from a 40 m high tower. How far from the base of the tower does the object land? | Position, Velocity, Acceleration | Deterministic | One | Define terms Solve the equation | Turn a verbal representation into an equation |
| D. Complex genetics problem | Using data from Cystic Fibrosis (CF) gene sequencing, restriction digest sites, and mutation probabilities, predict whether babies born to CF carrier parents will have CF or be carriers, and propose how to explain your prediction to parents. | Alleles, Chromosome segregation, Probabilities, Use of restriction digest sites for analysis of DNA sequences, Output of molecular analysis | Stochastic Probabilistic Emergent: analysis of molecular data | Many | All those in "A" | Analyze relationships Reduce and filter information Refine ambiguous goal states Synthesize data Evaluate evidence Argue from evidence Reflect on goal state and progress |
| E. Complex ecology problem | Devise a management strategy among multiple stakeholders for a small urban watershed to maximize water diversion and catchment, recreation, species preservation, and water quality. | Multiple stakeholders, Biodiversity, Water quality metrics, Catchment volume | Dynamic Emergent: intra- and inter-annual rainfall variation precipitation duration, timing, intervals, species responses, stakeholder investments | Many | All those in "B" | Analyze relationships Reduce and filter information Refine ambiguous goal states Synthesize data Evaluate evidence Argue from evidence Reflect on goal state and progress |
| F. Complex physics problem | Design a catapult that ejects a watermelon such that it passes through the uprights of a field goal post. | Catapult and elastic, Watermelon, Air/wind variables | Deterministic | Many | All those in "C" | Build a model Evaluate evidence, Argue from evidence Reflect on goal state and progress |